\documentclass{article}
\usepackage{amsmath, amssymb, amsfonts}
\title{Comments on : Frame Dragging Anomalies for Rotating Bodies}
\author{Hristu Culetu, \\Ovidius University, Dept.of Physics, \\B-dul Mamaia 124, 8700 Constanta, Romania, \\e-mail : hculetu@yahoo.com}

\begin{document}
\numberwithin{equation}{section}
\pagenumbering{arabic}
\maketitle
\newcommand{\fv}{\boldsymbol{f}}
\newcommand{\tv}{\boldsymbol{t}}
\newcommand{\gv}{\boldsymbol{g}}
\newcommand{\OV}{\boldsymbol{O}}
\newcommand{\wv}{\boldsymbol{w}}
\newcommand{\WV}{\boldsymbol{W}}
\newcommand{\NV}{\boldsymbol{N}}
\newcommand{\hv}{\boldsymbol{h}}
\newcommand{\yv}{\boldsymbol{y}}
\newcommand{\RE}{\textrm{Re}}
\newcommand{\IM}{\textrm{Im}}
\newcommand{\rot}{\textrm{rot}}
\newcommand{\dv}{\boldsymbol{d}}
\newcommand{\grad}{\textrm{grad}}
\newcommand{\Tr}{\textrm{Tr}}
\newcommand{\ua}{\uparrow}
\newcommand{\da}{\downarrow}
\newcommand{\ct}{\textrm{const}}
\newcommand{\xv}{\boldsymbol{x}}
\newcommand{\mv}{\boldsymbol{m}}
\newcommand{\rv}{\boldsymbol{r}}
\newcommand{\kv}{\boldsymbol{k}}
\newcommand{\VE}{\boldsymbol{V}}
\newcommand{\sv}{\boldsymbol{s}}
\newcommand{\RV}{\boldsymbol{R}}
\newcommand{\pv}{\boldsymbol{p}}
\newcommand{\PV}{\boldsymbol{P}}
\newcommand{\EV}{\boldsymbol{E}}
\newcommand{\DV}{\boldsymbol{D}}
\newcommand{\BV}{\boldsymbol{B}}
\newcommand{\HV}{\boldsymbol{H}}
\newcommand{\MV}{\boldsymbol{M}}
\newcommand{\be}{\begin{equation}}
\newcommand{\ee}{\end{equation}}
\newcommand{\ba}{\begin{eqnarray}}
\newcommand{\ea}{\end{eqnarray}}
\newcommand{\bq}{\begin{eqnarray*}}
\newcommand{\eq}{\end{eqnarray*}}
\newcommand{\pa}{\partial}
\newcommand{\f}{\frac}
\newcommand{\FV}{\boldsymbol{F}}
\newcommand{\ve}{\boldsymbol{v}}
\newcommand{\AV}{\boldsymbol{A}}
\newcommand{\jv}{\boldsymbol{j}}
\newcommand{\LV}{\boldsymbol{L}}
\newcommand{\SV}{\boldsymbol{S}}
\newcommand{\av}{\boldsymbol{a}}
\newcommand{\qv}{\boldsymbol{q}}
\newcommand{\QV}{\boldsymbol{Q}}
\newcommand{\ev}{\boldsymbol{e}}
\newcommand{\uv}{\boldsymbol{u}}
\newcommand{\KV}{\boldsymbol{K}}
\newcommand{\ro}{\boldsymbol{\rho}}
\newcommand{\si}{\boldsymbol{\sigma}}
\newcommand{\thv}{\boldsymbol{\theta}}
\newcommand{\bv}{\boldsymbol{b}}
\newcommand{\JV}{\boldsymbol{J}}
\newcommand{\nv}{\boldsymbol{n}}
\newcommand{\lv}{\boldsymbol{l}}
\newcommand{\om}{\boldsymbol{\omega}}
\newcommand{\Om}{\boldsymbol{\Omega}}
\newcommand{\Piv}{\boldsymbol{\Pi}}
\newcommand{\UV}{\boldsymbol{U}}
\newcommand{\iv}{\boldsymbol{i}}
\newcommand{\nuv}{\boldsymbol{\nu}}
\newcommand{\muv}{\boldsymbol{\mu}}
\newcommand{\lm}{\boldsymbol{\lambda}}
\newcommand{\Lm}{\boldsymbol{\Lambda}}
\newcommand{\opsi}{\overline{\psi}}
\renewcommand{\tan}{\textrm{tg}}
\renewcommand{\cot}{\textrm{ctg}}
\renewcommand{\sinh}{\textrm{sh}}
\renewcommand{\cosh}{\textrm{ch}}
\renewcommand{\tanh}{\textrm{th}}
\renewcommand{\coth}{\textrm{cth}}

\begin{abstract}
  It is shown that Collas and Klein ( ArXiv : 0811.2471 [gr-qc] ) wrongly concluded that ''negative frame dragging'' phenomenon takes place at all finite $r$ and $z$ coordinate values . We argue that a test particle with zero angular momentum counter-rotates with respect to the source in the ''time machine'' region only. In addition, Bonnor's spacetime has an event horizon at $r_{H}$ = 0.
\end{abstract}

 Keywords : event horizon, time machine region, frame dragging.\\

  These Comments concern the anomalous ''negative frame dragging'' phenomenon which appears, according to Collas and Klein \cite {CK1}, whenever\\
  ''zero angular momentum test particles acquire angular velocity in the opposite direction of rotation from the source of the metric''.
  
  We argue that the ''Proposition 1'' of pp. 4 is partially incorrect.\\
  Collas and Klein state that ''...~$\omega \prec 0 $ everywhere in $S_{B0}$'', probably on the grounds that $L \succ 0$ in their Eq. (7) leads to $\omega \prec 0$. But that is valid only when $M$ (or $n ) \succ 0$, where $n$ is given in Eq. (9). But why $n$ must be positive ? In authors' opinion, $h$ (with dimension of length squared) is a parameter related to rotation. We know there are two directions of rotation and therefore $h$ may be negative, too. 
  
  In the revised version \cite{CK2}, Collas and Klein justify their choice, $h \succ 0$, stating that ''we assume, without loss of generality, as in \cite{BS}, that $h \succ 0$ ''. But exactly the sign of $h$ (or of $M$) leads to the so called ''negative frame dragging'' effect. Therefore, we consider it is not a physical phenomenon produced by the choice of the sign of $h$. \\
  Let us notice that, when we pass from the Minkowski spacetime
\begin {equation}
ds^{2} = - dT^{2} + dR^{2} + dZ^{2} + R^{2} d\Phi^{2} 
\label{1}
\end{equation}
to the uniformly rotating one \cite{DS} 
\begin{equation}
ds^{2} = - ( 1 - \Omega^{2} R'^{2} ) dT'^{2} + dR'^{2} + dZ'^{2} + 2 \Omega R'^{2} d\Phi' dT' + R'^{2} d\Phi'^{2} 
\label{2}
\end{equation}
by means of the coordinate transformation
\begin{equation}
\Phi' = \Phi - \Omega T, ~~~~T' = T,~~~~Z' = Z,~~~~R' = R,
\label{3}
\end{equation}
the sign of the metric coefficient $g_{\Phi' T'} $ changes when the direction of rotation is reversed.\\ 
A similar effect takes place on $M$ in Eq. (1) of Ref. [1] : it could have both signs. Therefore, in our opinion, we have $\omega \prec 0$ only when $r \prec n$ (the ''time machine '' region), where closed timelike curves (CTC) are possible. \\
  In fact, even the authors of \cite{CK1} recognize at pp. 6, at the end of Chap. 3, that ''the sign of the metric coefficient $L$ determines the sign of the frame dragging $\omega$''. In other words, $L \prec 0$ (or  $r \prec \sqrt{2 |h|} $) leads to $\omega \prec 0$ and not $M$.   
  Similar conclusions were reached in \cite{HC1}. If we divide the two relations from Eq. (5.5) of Ref. [5] (with $\omega = 0)$, one obtains
   \begin{equation}
  \frac{\dot{\phi}}{\dot{t}} = \frac{d \phi}{d t} = \frac{L + bE}{(r^{2} - b^{2}) E - bL}
  \label{4}
  \end{equation}
  i.e. exactly Eq. (6) of Ref. [1], with $F = 1$, $L$ instead of $p_{\phi}$ and $(-b)$  instead of $M$ . Taking above a zero angular momentum particles , one get 
  \begin{equation}
  \frac{d \phi}{d t} = \frac{b}{r^{2} - b^{2}}
  \label{5}
  \end{equation}
  Here $b$ is considered to be positive since its sign depends upon how we define the ''improper'' time translation in Eq. (2.2), Ref. [5]. In conclusion, in our view, the negative value of $\omega$ in (7), Ref. [1] has nothing to do with region $S_{B0}$ but comes from the negative value of $g_{\phi \phi}$ (the time machine region $r \prec b$ in \cite{HC1}). Its boundary $r = b$ is the velocity of light surface. Because $g_{\phi t} \neq 0$ when $g_{\phi \phi} = 0$, the metric  (2.5) (with $\omega = 0$) from \cite{HC1} is nonsingular at $r = b$. Therefore, the timelike curves may cross into the time machine region and viceversa \cite{CGLP}. 
  
  One should finally mention the problem of the existence of an event horizon in Bonnor's dust metric. Collas and Klein argued in Chap.4 (''Concluding remarks'' of \cite{CK3}) that :\\
  ''The spacetime considered here has some unrealistic features. It has an isolated singularity with no event horizon''.\\
  But their spacetime (1) of \cite{CK3} (and even the metric (1) of Ref. [1]) has an event horizon which is obtained from
  \begin{equation}
  g_{tt} - \frac{g_{t \phi}^{2}}{g_{\phi \phi}} = 0,
  \label{6}
  \end{equation}
  (see Eq. (7) of \cite{HC2} or Eq. (7) of \cite{HC3}). Eq. (6) leads to $r_{H} = 0$, i.e. the horizon is located on the rotation axis, in the interior of the time machine region. The fact that the numerator from the l.h.s. of (6) equals $r^{2}$ represents the Collas and Klein ''coordinate condition'' (in \cite{CK1}) or ''gauge condition'' in \cite{CK3}).

\end{document}